\documentclass[graybox]{svmult}

\usepackage{mathptmx}       
\usepackage{helvet}         
\usepackage{courier}        
\usepackage{type1cm}        
\usepackage{makeidx}         
\usepackage{graphicx}        
\usepackage{multicol}        
\usepackage[bottom]{footmisc}
\usepackage{amsmath}
\usepackage{amssymb}
\usepackage[squaren]{SIunits} 
\usepackage[labelsep=space, labelfont=bf]{caption}
\usepackage{subcaption}
\let\myurl\url
\usepackage{hyperref}
\let\url\myurl
\usepackage{multirow}
\usepackage{xcolor}

\makeindex             

\newcommand{\statdom}{\Omega_\mathrm{s}}
\newcommand{\rotdom}{\Omega_\mathrm{r}}
\newcommand{\interface}{\Gamma_\mathrm{I}}
\newcommand{\boundrot}{\Gamma_\mathrm{r}}
\newcommand{\boundstat}{\Gamma_\mathrm{s}}
\newcommand{\THD}{\mathrm{THD}}
\newcommand{\Wmag}{W_{\mathrm{mag}}}
\begin{document}

\title*{Uncertainty Quantification For A Permanent Magnet Synchronous Machine With Dynamic Rotor Eccentricity}
\titlerunning{UQ for a PMSM with dynamic rotor eccentricity}
\author{Zeger Bontinck, Oliver Lass, Herbert De Gersem and Sebastian Sch\"ops}
\institute{Zeger Bontinck \at Technische Universit\"at Darmstadt, Graduate School of Computational Engineering,
Dolivostr. 15, 64293 Darmstadt, Germany, \email{bontinck@gsc.tu-darmstadt.de}
\and Oliver Lass \at  Technische Universit\"at Darmstadt, Department of Mathematics, Chair of Nonlinear Optimization, Dolivostr. 15,
64293 Darmstad, Germany, \email{lass@mathematik.tu-darmstadt.de}
\and Herbert De Gersem \at Technische Universit\"at Darmstadt, Institut f\"ur Theorie Elektromagnetischer Felder, Schlossgartenstr. 8, 64289 Darmstadt, Germany, \email{degersem@temf.tu-darmstadt.de}
\and Sebastian Sch\"ops \at Technische Universit\"at Darmstadt, Graduate School of Computational Engineering,
Dolivostr. 15, 64293 Darmstadt, Germany, \email{schoeps@gsc.tu-darmstadt.de}}
\maketitle

\abstract*{The influence of dynamic eccentricity on the harmonic spectrum of the torque of a permanent magnet synchronous machine is studied. The spectrum is calculated by an energy balance method. Uncertainty quantification is applied by using generalized Polynomial Chaos and Monte Carlo. It is found that the displacement of the rotor impacts the spectrum of the torque the most.}

\abstract{The influence of dynamic eccentricity on the harmonic spectrum of the torque of a permanent magnet synchronous machine is studied. The spectrum is calculated by an energy balance method. Uncertainty quantification is applied by using generalized Polynomial Chaos and Monte Carlo. It is found that the displacement of the rotor impacts the spectrum of the torque the most.}

\section{Introduction}
\label{sec:intro}
During mass production of permanent magnet synchronous machines (PMSMs) small deviations on the machine are introduced. These deviations may lead to underperformance. Therefore the field of uncertain quantification is gaining more and more interest for the machine simulation. The rotor position is a typical uncertainty in an electrical machine design. It is known that rotor eccentricity increases the cogging torque (see e.g.~\cite{Coenen_2011aa}). In the literature different types of eccentricity are identified. Static eccentricity means that the rotor is shifted out of the center of the stator but still rotates around the center of the rotor. In the case of dynamic eccentricity, the rotor rotates around the center of the stator. Also more complex eccentricities like e.g. an inclined rotor shaft are possible (see e.g.~\cite{Bontinck_2016aa}). 
In this paper the influence of dynamic eccentricity on the torque and its total harmonic distortion will be studied. In~\cite{Sadowski_1992aa} an overview and comparison of different methods for torque calculation can be found. The most used method relies on the Maxwell stress tensor. This method, however, is very sensitive to the applied space discretization and to the location in the airgap where the quantities are evaluated. As a consequence big numerical deviations are easily introduced \cite{Mizia_1988aa, Tarnhuvud_1988aa}. Another way to calculate the torque is to use the energy balance method, which relies on the principle of conservation of energy (see e.g. \cite{Silwal_2014aa}).

The introduction of uncertainties in the PMSM model is the first step in a bigger procedure in which the magnets of the machine are subjected to robust optimization. This can be performed by analyzing the worst-case scenario or by taking into account the standard deviation. In both cases, an accurate and reliable representations of rotor eccentricity is needed. Moreover, a fast simulation technique for calculating the according statistic moment of the harmonics of the torque is indispensable.

The paper is structured as follows: in Sect. \ref{sec:prob} the PMSM is introduced. It is explained how the torque is calculated and how the eccentricity is modeled. In Sect. \ref{sec:results}  the results are presented and discussed and finally conclusions are drawn.

\section{Problem Description}
\label{sec:prob}
\begin{figure}[b]      
\centering
		\subcaptionbox{\label{fig:mach1p}Detailed view of the PMSM. The blue dashed line indicates the location of the interface. The magnets are marked in gray.}[0.47\textwidth]{ \def\svgwidth{0.4\textwidth}
		      \input{geometry_machine2.pdf_tex}}
		      \hspace{1em}
		\subcaptionbox{\label{fig:dyn}Schematic representation of the full machine with rotor eccentricity (figure adapted from \cite{Belmans_1987ab}).}[0.49\textwidth]{     \def\svgwidth{0.48\textwidth}
		      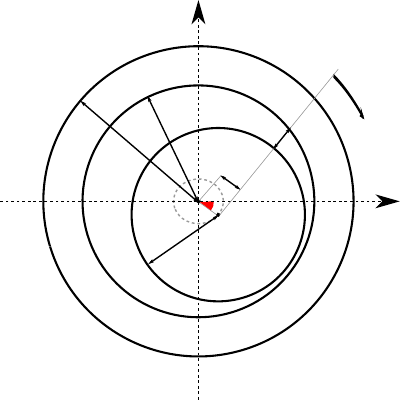}

\caption{\label{fig:mach}Crosssectional view of the PMSM submitted to uncertainty quantification}
\end{figure}

The PMSM has six slots per pole and a double layered winding scheme with two slots per pole per phase (Fig.~\ref{fig:mach1p}). The machine has length $\ell_z = \unit{10}{mm}$. The machine is constructed of laminated steel with a relative permeability $\mu_r=500$.
Let $\statdom$ and  $\rotdom$ depict the stator and rotor domain respectively. Define $\interface=\partial\statdom \cap \partial\rotdom$, $\boundstat=\partial\statdom\setminus\interface$ and $\boundrot=\partial\rotdom\setminus\interface$. To calculate the torque of the machine the magnetostatic approximation of the Maxwell's equations has to be solved for both domains. This implies that the eddy and displacement currents are neglected and one obtains the semi-elliptic partial differential equations 
\begin{subequations}
\begin{equation}\label{eq:mqscont_stat}
    \vec{\nabla}\times\left(\nu\vec{\nabla}\times\vec{A}_{\mathrm{s}}(t)\right) =\vec{J}_{\rm src}(t), \quad \mbox{on } \statdom
\end{equation}
\begin{equation}\label{eq:mqscont_rot}
    \vec{\nabla}\times\left(\nu\vec{\nabla}\times\vec{A}_{\mathrm{r}}(t)\right) =-\nabla\times\left(\nu\mathbf{B}_{\rm rem}\right), \quad \mbox{on } \rotdom
\end{equation}
with Dirichlet boundary conditions $\vec{A}_{\mathrm{s}}|_{\boundstat} =0$ and $\vec{A}_{\mathrm{r}}|_{\boundrot} =0$. At the interface it holds that
\begin{equation}\label{eq:interface}
    \nu\vec{n}\times \vec{A}_{\mathrm{s}}|_{\interface}\left(\vec{Y}(t)\right)-\nu\vec{n}\times \vec{A}_{\mathrm{r}}|_{\interface}\left(\vec{Y}(t)\right)=0,
\end{equation}
\end{subequations}
  where $\vec{Y}(t)$ describes how the interfaces are connected at every time step~$t$. The reluctivity is depicted by a scalar $\nu$ since only linear isotropic materials are considered. $\vec{A}(t)$ is the magnetic vector potential, $\vec{J}_{\rm src}(t)$ represents the imposed source current density, which is related to the applied currents in the coils, and $\mathbf{B}_{\rm rem}$ the remanence of the permanent magnets. The applied current density is aligned with the $z$-direction, whereas the remanence is in the $xy$-plane. It is generally accepted that machines are adequately modeled in 2D, meaning that the magnetic field has no $z$-component: $\vec{B}(t)=(B_x(t), B_y(t), 0)$. Since $\vec{B}(t)=\nabla\times\vec{A}(t)$, one can write $\vec{A}(t) = (0, 0, A_z(t))$ and also $\vec{J}$ has only a $z$-component. Discretizing $\vec{A}_{\mathrm{s}}$ by linear edge shape functions $\vec{w}_i(x,y)$ one approximates
\begin{equation}
\vec{A}_{\mathrm{s}}\approx\sum_{i=1}^N a^{\mathrm{(s)}}_i\vec{w}_i(x,y)=\sum_{i=1}^N a^{\mathrm{(s)}}_i\frac{N_i(x,y)}{\ell_z}\vec{e}_z,
\end{equation}
where $N_i(x,y)$ depicts the nodal finite elements which are associated with the triangulation of the machine's cross-section and $\vec{e}_z$ is the unit vector in $z$-direction. Using the Ritz-Galerkin approach one finds the discretized form 

\begin{equation}
\label{eq:discr_stat}
\mathbf{K}^{\mathrm{(s)}}_\nu \mathbf{a}^{\mathrm{(s)}}(t)+\mathbf{j}_{\rm int}^{\mathrm{(s)}} =\mathbf{j}_{\mathrm{src}}(t)\quad \text{and}\quad\mathbf{K}^{\mathrm{(r)}}_\nu \mathbf{a}^{\mathrm{(r)}}(t)+\mathbf{j}_{\rm int}^{\mathrm{(r)}} =\mathbf{j}_{\mathrm{pm}},
\end{equation}
where $\mathbf{K}^{\mathrm{(s)}}_\nu$ and $\mathbf{K}^{\mathrm{(r)}}_\nu$ are the finite element system matrices for the stator and rotor, respectively, $\mathbf{a}^{\mathrm{(s)}}$ and $\mathbf{a}^{\mathrm{(r)}}$ depict the degrees of freedom (DoFs) and $\mathbf{j}_{\mathrm{src}}$ ($\mathbf{j}_{\mathrm{pm}}$) are the discretized versions of the current densities (permanent magnets). $\mathbf{j}_{\rm int}^{\mathrm{(s)}}$ and $\mathbf{j}_{\rm int}^{\mathrm{(r)}}$ are the discretized versions of the conditions at $\interface$.

A time-stepping technique based on \cite{Preston_1988aa} is implemented for the rotation of the machine. In the airgap a contour is defined which splits the full domain in two parts: a fixed outer domain connected to the stator $\statdom$ and an inner domain connected to the rotor $\rotdom$, where the mesh will be rotated. At the contour the nodes are distributed equidistantly. The time step is chosen so that the inner domain is rotated by exactly three nodes between successive time steps. The connections between the nodes on the contour and the ones in the airgap are updated at every time step. One can write $\mathbf{j}_{\rm int}^{(s)}=\mathbf{P}_s\lambda$ and $\mathbf{j}_{\rm int}^{(r)}=\mathbf{P}_r(t)\lambda$ with $\lambda$ the Lagrange multiplier and projectors $\mathbf{P}{_\mathrm{r}}$ and $\mathbf{P}{_\mathrm{s}}\in\{-1,0,1\}^N$. This yields 

\begin{equation}
\label{eq:discr}
\underbrace{\begin{bmatrix}
    \mathbf{K}^{\mathrm{(s)}}_\nu       & 0 & \mathbf{P}{_\mathrm{s}} \\
    0       & \mathbf{K}^{\mathrm{(r)}}_\nu & \mathbf{P}{_\mathrm{r}}(t) \\
    \mathbf{P}{_\mathrm{s}}^\top       & \mathbf{P}{_\mathrm{r}}^\top (t)& 0
\end{bmatrix}}_{\mathbf{K}_\nu(t)}
\begin{bmatrix}
    \mathbf{a}^{\mathrm{(s)}} \\
    \mathbf{a}^{\mathrm{(r)}}\\
    \lambda
\end{bmatrix} 
=
\begin{bmatrix}
    \mathbf{j}_{\mathrm{src}}(t) \\
    \mathbf{j}_{\mathrm{pm}} \\
    0
\end{bmatrix} 
\end{equation}
with $\mathbf{K}_{\nu}(t)$ the finite element system matrix \cite{Salon_1995aa}. Solving for $\mathbf{a}^{(\mathrm{s})}$ and $\mathbf{a}^{(\mathrm{r})}$  enables the calculation of the torque by using an energy balance method. The change in the stored magnetic energy can be written as
\begin{equation}
\label{eq:EBM}
\frac{\mathrm{d}\Wmag(t)}{\mathrm{d}t}=P_{\text{e}}(t)-P_{\text{l}}(t)-P_{\text{m}}(t),
\end{equation}
with $P_{\text{e}}$ the electrical energy, the mechanical energy $P_{\text{m}}$ and the losses $P_{\text{l}}$. To calculate the torque a time-averaging approach is used, i.e. integrating Eq. \eqref{eq:EBM} over one period $T$. Due to the conservation of energy $\Wmag(t)=\Wmag(t+T)$ and one finds that $\bar{P}_{\text{m}}=\bar{P}_{\text{e}}-\bar{P}_{\text{l}}$. The total electrical energy is given by
\begin{equation}
\bar{P}_{\text{e}}=\frac{1}{T}\int_0^T P_{\text{e}}(t) \; \text{d}t=\frac{1}{T}\int_0^T \mathbf{u}^\top_{\text{str}}(t)\mathbf{i}_{\text{str}}(t) \; \text{d}t, 
\end{equation}
where  $\mathbf{i}_{\text{str}}$ depicts the current in the coils, such that $\mathbf{j}_{\mathrm{src}}=\mathbf{X}_{\text{str}}\mathbf{i}_{\text{str}}$, with $\mathbf{X}_{\text{str}}$ the winding function \cite{Schops_2013aa}. The voltages of the stranded conductors $\mathbf{u}_{\mathrm{str}}$ are calculated from the solutions of Eq.~\eqref{eq:discr} by
\begin{equation}
\label{eq:str_vol}
\mathbf{u}_{\text{str}}(t)=\mathbf{R}_{\mathrm{str}}\mathbf{i}_{\text{str}}(t)+\frac{\text{d}}{\text{d}t} (\mathbf{X}^{\top}_{\text{str}}\mathbf{a}(t)),
\end{equation}
with $\mathbf{R}_{\text{str}}$, the DC resistance of coils. The losses are given by
\begin{equation}
\bar{P}_{\text{l}}=\frac{1}{T}\int_0^T P_{\text{l}}(t) \; \text{d}t =\frac{1}{T}\int_0^T  \mathbf{i}^\top_{\text{str}}(t)\mathbf{R}_{\text{str}}\mathbf{i}_{\text{str}}(t) \; \text{d}t. 
\end{equation}
For the time-averaged mechanical energy one finds
\begin{equation}
\bar{P}_{\text{m}}= \frac{1}{T}\int_0^T P_{\text{m}}(t) \; \text{d}t= \frac{1}{T} \int_0^T\left(P_\text{e}(t)-P_\text{l}(t)\right)\; \text{d}t
\end{equation} leading to time-averaged torque 
\begin{equation} 
\bar{\tau}_\text{0}=\frac{1}{T}\int_0^T \tau_{\text{m}}(t) \; \text{d}t:=\frac{1}{T\omega_{\text{m}}}\int_0^T P_{\text{m}}(t) \; \text{d}t.
\end{equation} The mechanical angular frequency is $\omega_\text{m}=\omega_\text{e}/N_p$, with $\omega_\text{e}$ the electric angular frequency and $N_p=6$ the pole pair number. To calculate the higher harmonics of the torque, Eq.~\eqref{eq:EBM} is solved, meaning
\begin{equation}
\tau_{\text{m}}(t) =\frac{1}{\omega_\text{m}}\left( P_{\text{e}}(t)-P_{\text{l}}(t)-\frac{\mathrm{d}\Wmag(t)}{\mathrm{d}t}\right),
\end{equation}
with $\Wmag=1/2\mathbf{a}^\top \mathbf{K}_\nu\mathbf{a}$, where $\mathbf{a}$ depicts all DoFs. Taking a Fourier transform of $\tau_{\text{m}}$ one can then define the total harmonic distortion (THD) as 
\begin{equation}
\THD=\frac{\sum_{i=1}^n \tau_i}{\tau_\text{0}},
\end{equation}
 where $\tau_i$ depicts the harmonics of the torque. 

The eccentricity of the rotor is described by two parameters: $R_0(\omega)$, which defines the initial displacement from the central position, and $\theta_0(\omega)$, which depicts the direction of its displacement (see Fig.~\ref{fig:dyn}). The stochastic nature of a quantity is indicated by $\omega$. It is assumed that $R_0,\theta_0 \in L^2(\Theta,\Sigma,P)$ are independent random variables from the probability space $(\Theta,\Sigma,P)$ and that $R_0$ is normal distributed and $\theta_0$ uniformly distributed, $R_0\sim\mathcal{N}(0,\sigma_{R_0}^2)$ and $\theta_0\sim\mathcal{U}(0,\pi)$, with $\sigma_{R_0}=\unit{0.4/3}{mm}$, which corresponds to an eccentricity of $13\%$. 

\begin{figure}[t]
\centering
\subcaptionbox{\label{fig:spectrum} Part of the spectrum for a nominal machine and for a machine with $\varepsilon=\unit{50}{\%}$.}[0.46\textwidth]{\includegraphics[width=0.47\textwidth]{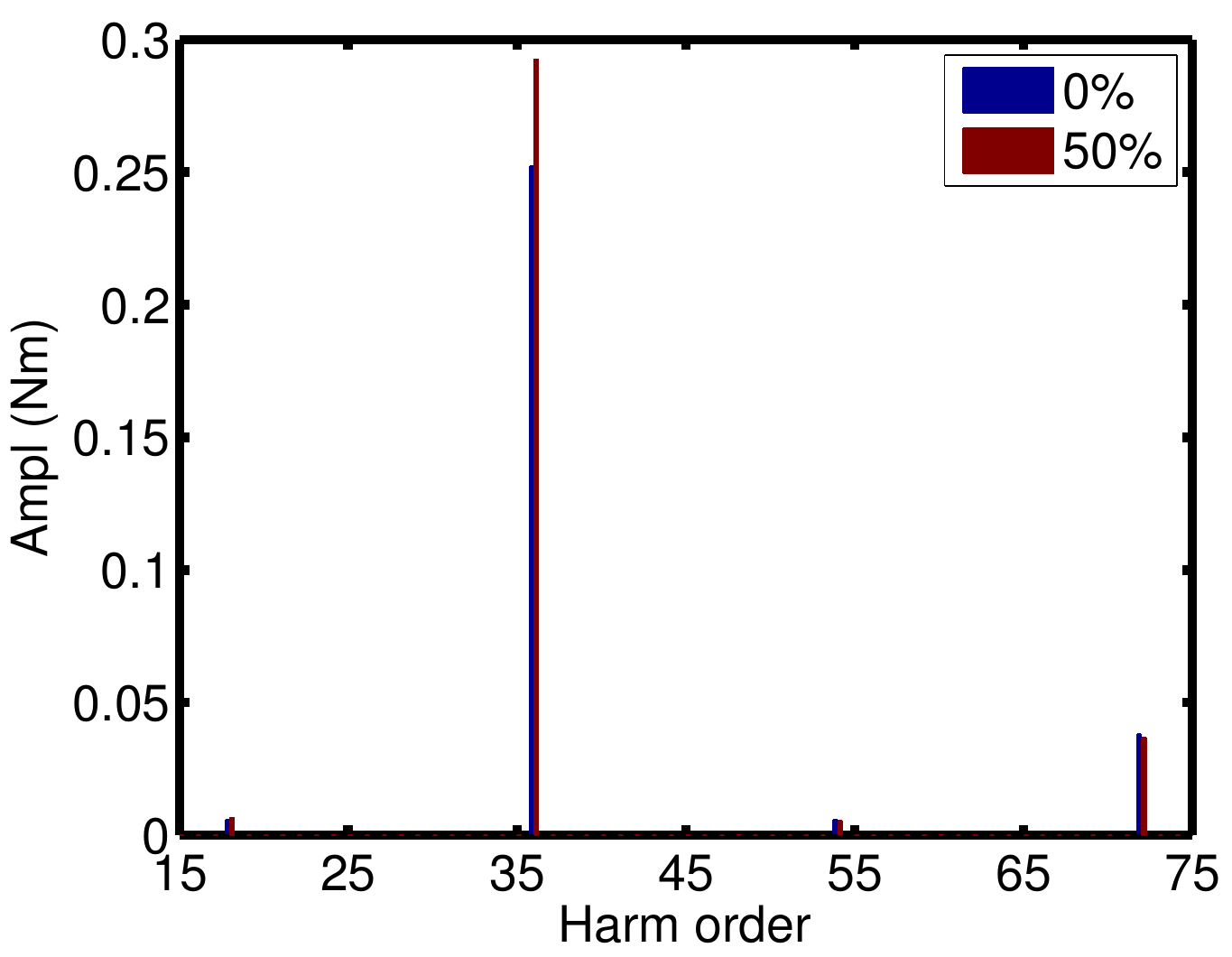}}
\hspace{0.5em}
\subcaptionbox{\label{fig:THD}The total harmonic distortion as a function of the rotor eccentricity.}[0.46\textwidth]{\includegraphics[width=0.45\textwidth]{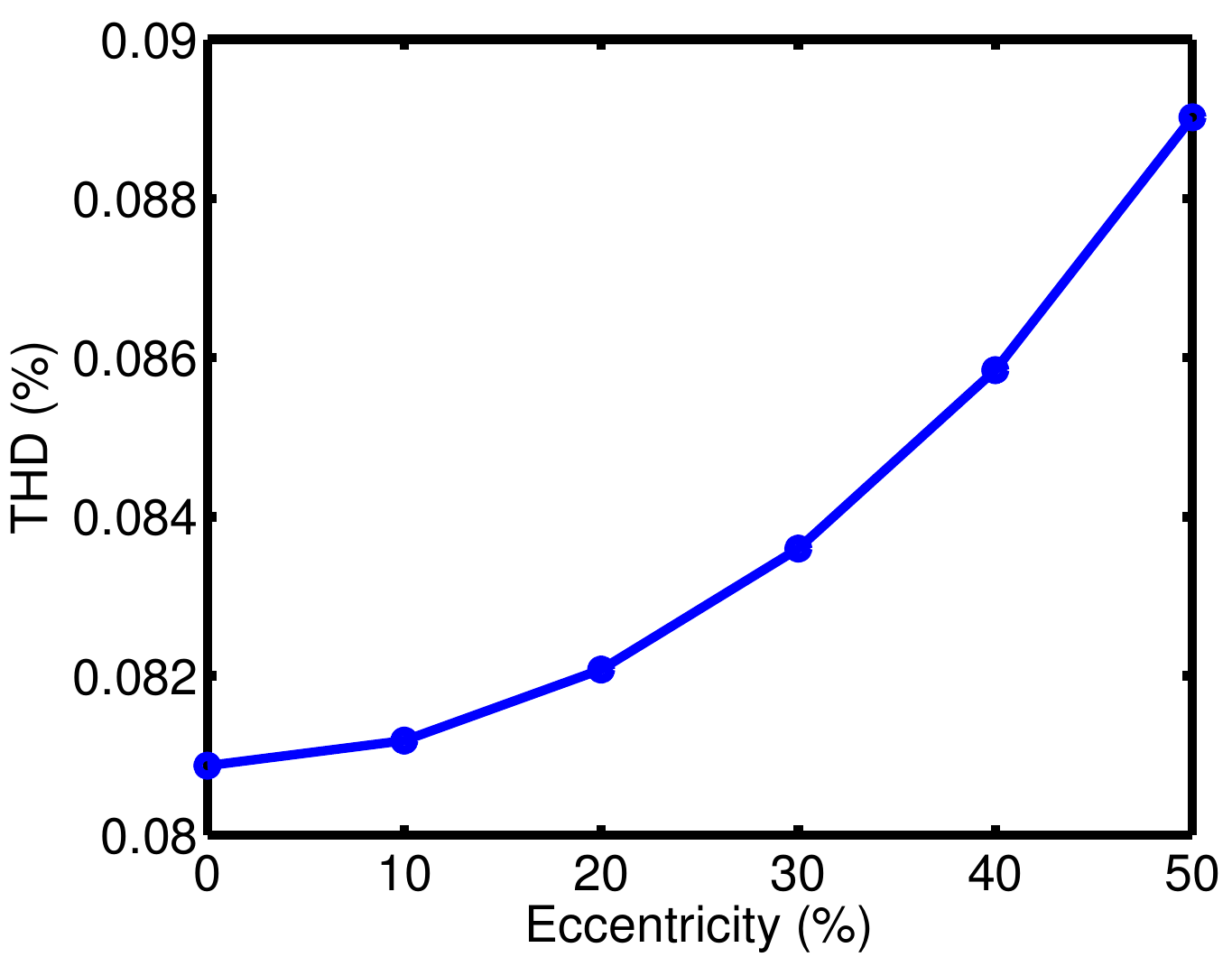}}
\caption{\label{fig:harmonics}Study of the influence of the eccentricity on the torque spectrum}
\end{figure}

Due to the chosen approach for modeling the rotation, the eccentricity is implicitly considered according to the dynamic model of \cite{Frohne_1968aa}, meaning that the airgap width $\delta(x,t)$ (see Fig.~\ref{fig:dyn}) can be expressed as a function of the arc $x$, such that
\begin{equation}
\label{eq:dyn_ecc}
\delta(x,t)=\delta_\text{m}\left[1-\varepsilon_\text{m}(\omega)\cos(\alpha(x,t,\omega))\right],
\end{equation}
with the eccentricity $\varepsilon_\text{m}(\omega)= R_0(\omega)/\delta_\text{m}=R_0(\omega)/(R_{\text{st,in}}-R_{\text{rt}})$, where $\delta_\text{m}$ is the mean mechanical airgap width. The angle $\alpha$ is given by
\begin{equation}
\label{eq:alpha}
\alpha(x,t,\omega)=x-\omega_\text{m} t-\theta_0(\omega).
\end{equation}
Fig.~\ref{fig:harmonics} illustrates the influence of the eccentricity on the spectrum of the torque. In Fig.~\ref{fig:spectrum} the spectra for a nominal machine and for a machine with $\varepsilon_\text{m}=\unit{50}{\%}$ eccentricity are shown. The 36th harmonic, which is related to the cogging torque, is increased due to the eccentricity. To study the full effect of eccentricity on the spectrum, the THD is calculated. One sees in Fig.~\ref{fig:THD} that this quantity is increasing with a higher eccentricity. This implies that an eccentric machine suffers from increased undesired effects, e.g. \cite{Dorrell_2005aa}.

Let $Z(\omega)=f(R_0(\omega),\theta_0(\omega))$ be a discrete random variable then the expectation value and the variance can be approximated by
\begin{equation}
 \label{eq:expv_tau}
 \mu_{Z}=\mathbb{E}[Z]\approx\sum_{j=1}^{N}w_jf_j \quad \mathrm{and}\quad \sigma^2_{Z}=\mathbb{V}[Z]\approx\sum_{j=1}^{N}w_j\left(f_j-\mu_z\right)^2,
\end{equation}
respectively, where $f_j=f(R_0^{(j)},\theta_0^{(j)})$ denotes a random outcome. The choice of weights $w_j$ and samples $f_j$ depends on the method. In the Monte Carlo (MC) approach the samples are chosen randomly and have an equal weight $1/N$. For collocation based on generalized Polynomial Chaos (gPC) they are determined according to quadrature rules. The gPC bases for normal and uniform distributions are constructed from Hermite polynomials and Legendre polynomials, respectively. In this paper we use both methods for the uncertainty quantification. With aid of the MC sampling a variance based sensitivity analysis is performed based on \cite{Saltelli_2010aa}. However, it is known, that for a low number of random variables gPC converges faster \cite{Xiu_2010aa} than Monte Carlo (see e.g. \cite{Dick_2013aa}). To avoid numerical noise, remeshing for every sample $j$ is avoided. Instead, the original mesh is mapped to a new one as in~\cite{Bontinck_2016aa}.

\section{Results}
\label{sec:results}
The nominal machine, i.e. $R_0=\unit{0}{mm}$ and $\theta_0=0$ has a torque $\tau_\text{0}=\unit{4.060}{Nm}$ and a $\THD=\unit{0.809}{\%}$. The UQ was performed by using gPC with $5\times5$ tensor product grid and by using $N_\text{MC}=3200$ MC samples for the uncertain input parameters. For $\mu_\tau$ a value of $\unit{4.066}{Nm}$ is retrieved by both approaches. The found standard deviation is $\unit{8.1\cdot10^{-3}}{Nm}$. For the THD one finds $\mu_\THD=\unit{0.0814}{\%}$ and $\sigma_\THD=\unit{7.8\cdot10^{-4}}{\%}$. The error $\epsilon_{\mathrm{MC}}$ on $\mu$ for MC is estimated by $\sigma_{Z}/ \sqrt{N_\text{MC}}$. For the torque one finds $\epsilon_{\mathrm{MC}}=\unit{1.5\cdot10^{-4}}{Nm}$ and for the THD $\epsilon_{\mathrm{MC}}=\unit{0.14\cdot10^{-4}}{\%}$ . The sensitivity of $R_0\approx 1$, where the sensitivity of $\theta_{0}<10^{-6}$. 

\section{Conclusions}
Uncertainty quantification was performed to determine the effects of rotor eccentricity on the torque and on the total harmonic distorting on the torque. It is found that the displacement, rather than the direction of the displacement influences these quantities the most.

\begin{acknowledgement}
This work is supported by the German BMBF in the context of the SIMUROM project (grant number 05M2013), by the ‘Excellence Initiative’ of the German Federal and State Governments and the Graduate School of Computational Engineering at TU Darmstadt.
\end{acknowledgement}

\end{document}